\documentstyle[12pt]{article}
\begin{document}
\newcommand{\be}{\begin{equation}}
\newcommand{\ee}{\end{equation}}
\newcommand{\ba}{\begin{eqnarray}}
\newcommand{\ea}{\end{eqnarray}}
\newcommand{\p}{\partial}
\newcommand{\fr}{\frac}

\begin{center}
{\bf S.S. Gershtein, A.A. Logunov, and  M.A. Mestvirishvili}

\vspace*{0.5cm}
{\bf UPPER LIMIT ON THE GRAVITON MASS}
\end{center}

\vspace*{0.5cm}
The problem of the existence of nonzero invariant mass of the graviton 
can have the fundamental significance. The estimate for the upper bound 
on the graviton mass $(m_g<2\cdot 10^{-62}\mbox{г})$ has been derived
in Ref. [1], where the authors used the data on the existence of the 
gravitational coupling between the galaxy clusters, which is not cut off 
by the Yukawa potential at least up to distances $\sim$500~Kpc. 
In this paper we will present the estimates for upper limit on the
graviton mass basing on the observable parameters of the Universe expansion.
The fact that in this case typical distances are 3-4 orders of magnitude larger than
those between gravitationally bound galaxy clusters allows one to 
strengthen the estimates on the upper limit on the graviton mass by
few orders of magnitude, respectively.

It should be noted that introduction of the nonzero invariant mass of the
graviton requires to go beyond the General Theory of Relativity (GTR). 
This can be done naturally by using the notions of the gravitational
field in the Minkowsky space [2,3]. 
In Ref. [3] the complete energy-momentum tensor
$t^{\mu\nu}$ (including the gravitational field), which is conserved 
in the Minkowsky space, is considered as a source 
of the gravitational field described
by symmetric tensor $\Phi^{\mu\nu}$.
In arbitrary fixed (not necessarily inertial) frame of the Minkowsky
space with metric tensor $\gamma_{\mu\nu}$ the equations for
the density of the gravitational field $\tilde\Phi^{\mu\nu}$
can be written analogously to the Maxwell equations and Lorentz condition
for the electromagnetic field as follows:
\be
(\gamma^{\alpha\beta}D_\alpha D_\beta+m^2_g)\tilde\Phi^{\mu\nu}=16\pi \tilde t^{\mu\nu},
\ee
\be
D_\nu \tilde\Phi^{\mu\nu}=0\;,
\ee
where $D_\alpha$ is the covariant derivative in the Minkowsky space, 
$m_g$ is the graviton mass $(\hbar=c=G=1)$, and $\tilde\Phi^{\mu\nu}, 
\tilde t^{\mu\nu}$ are the densities of tensors:
\be
\tilde\Phi^{\mu\nu}=\sqrt{-\gamma}\Phi^{\mu\nu},\;\;\;
\tilde t^{\mu\nu}=\sqrt{-\gamma}t^{\mu\nu},\;\;\;
\gamma=\det (\gamma_{\mu\nu})=\det (\tilde\gamma^{\mu\nu})\;.
\ee
Condition (2) singles out polarization states with spin values of 2 and 0
and provides the conservation of the density of the energy-momentum
tensor $D_\mu\tilde t^{\mu\nu}=0$ in Eq. (1). 
The density of the energy-momentum tensor is defined, following Hilbert,
by Euler's variation with $\gamma_{\mu\nu}$ metric of the
Lagrangian density of the system
\be
\tilde L=\tilde L_g (\gamma_{\mu\nu}, \Phi_{\mu\nu})+
\tilde L_M (\gamma_{\mu\nu}, \Phi_{\mu\nu}, \Phi_A)\;,
\ee
where $\tilde L_g$ is the density of the gravitational field Lagrangian,
and $\tilde L_M$ corresponds to the density of Lagrangian of the matter
described by the $\Phi_A$ fields.
\be
\tilde t^{\mu\nu}=-2\frac{\delta\tilde L}{\delta\gamma_{\mu\nu}}\;,
\ee
where  Euler's variation is
\be
\frac{\delta\tilde L}{\delta\gamma_{\mu\nu}}=
\frac{\delta\tilde L}{\partial\gamma_{\mu\nu}}-\partial_\sigma
\left (
\frac{\partial\tilde L}{\partial\gamma_{\mu\nu,\sigma}}\right )\;;
\gamma_{\mu\nu,\sigma}=\frac{\partial\gamma_{\mu\nu}}{\partial x^\sigma}\;.
\ee
On can derive equations for the gravitational  field and matter fields 
from the least action principle
\be
\frac{\delta\tilde L}{\delta\tilde\Phi^{\mu\nu}}=0,\;\;\;
\frac{\delta\tilde L}{\delta\Phi_A}=0.
\ee
To get the form of (1) and (2) for these equations it is necessary
to have the density of the gravitational field
$\tilde\Phi^{\mu\nu}$ coming into the density of the matter Lagrangian
$\tilde L_M$ in combination with the density of the metric tensor 
$\tilde\gamma^{\mu\nu}$:
\be
\tilde g^{\mu\nu}=\tilde\gamma^{\mu\nu}+\tilde\Phi^{\mu\nu},\;\;
\tilde g^{\mu\nu}=\sqrt{-g}g^{\mu\nu},\;\;
g=\det (g_{\mu\nu})=\det (\tilde g^{\mu\nu})\;,
\ee
i.e. $\tilde L_M (\tilde g^{\mu\nu}, \Phi_A)$. It means that
the motion of the matter subjected to the gravitational filed
looks like as it could take place in the Riemann space with the metric
$g_{\mu\nu}$. The Lagrangian density resulting to Eqs. (1) and (2)
has the form
\be
\tilde L=\tilde L_g+\tilde L_M(\tilde g^{\mu\nu}, \Phi_A),
\ee
\be
\tilde L_g =\frac{1}{16\pi}\tilde g^{\mu\nu}
(G_{\mu\nu}^\lambda G^\sigma_{\lambda\sigma}-G^\lambda_{\mu\sigma}
G^\sigma_{\nu\lambda})-\frac{m^2}{16\pi}
\left (\frac{1}{2}\gamma_{\mu\nu}\tilde g^{\mu\nu}-\sqrt{-g}
-\sqrt{-\gamma}
\right )\;,
\ee
where the $G_{\mu\nu}^\lambda$ values are the components of the
\underline{tensor} 
\be
G_{\mu\nu}^\lambda=\frac{1}{2}g^{\lambda\sigma}
(D_\mu g_{\nu\sigma}+D_\nu g_{\mu\sigma}-D_\sigma g_{\mu\nu})\;,
\ee
and due to this fact the $\tilde L_g$ value behaves as the density of the 
scalar under \underline{any} coordinate transformations. Using (9) and
(10) and taking into account (7) one can write
the equations for the gravitational field in the form of [3]
\be
\left (R^\mu_\nu-\frac{1}{2}\delta^\mu_\nu R\right )+\frac{m^2_g}{2}
(\delta^\mu_\nu+g^{\mu\alpha}\gamma_{\alpha\nu}-\frac{1}{2}
\delta^\mu_\nu g^{\alpha\beta}\gamma_{\alpha\beta})=8\pi T^\mu_\nu
\ee
\be
D_\nu\tilde g^{\mu\nu}=0\;,
\ee
where $T^\mu_\nu$ is the matter energy-momentum tensor in the Riemann space.

From these equations one gets the equations for the matter
\be
\nabla_\nu \tilde T^{\mu\nu}=0,\;\;\;  \tilde 
T^{\mu\nu}=-2\frac{\delta \tilde L_M} {\delta g_{\mu\nu}}\;,
\ee
where $\nabla_\nu$ is the covariant derivative in the effective
Riemann space. Eqs. (12) and  (13) are covariant with respect to any 
coordinate transformations and form-invariant under Lorentz's transformations. 

Writing down the interval of the effective Riemann space for the homogeneous 
and isotropic Universe in the form of
\be
ds^2=U(t)dt^2-V(t)
\left [\frac{dr^2}{1-kr^2}+r^2(d\Theta^2+\sin^2\Theta d\Phi^2)\right ]\;,
\ee
(where $k=1,\; -1,\; 0$ for the closed, hyperbolic and ``flat'' Universe),
one gets from Eqs. (13) 
\be
\frac{\partial}{\partial t}\sqrt{\frac{V^3}{U}}=0,\;\;\;
\mbox{т.е.}\;\; V=aU^{1/3}, a=\mbox{const}.
\ee
\be
\frac{\partial}{\partial r}[r^2(1-kr^2)^{1/2}]-2r
(1-kr^2)^{-1/2}=0\;,
\ee
Eq. (17) is valid only for $k=0$. Thus, \underline{the Universe can be 
only ``flat''} (i.e. its  space geometry is Euclidian).
Using the proper time $d\tau=U^{1/2}dt$ and denoting $R^2=U^{1/3}$
one can write interval (15) in the form
\be
ds^2=d\tau^2-aR^2(\tau)(dx^2+dy^2+dz^2)\;.
\ee
In this case Eqs. (12) in inertial frame take the form\footnote{Note, that
the metric of the Minkowsky space $\gamma_{\mu\nu}$ comes into Eq. (12). 
Due to this fact the Minkowsky space becomes observable, and 
the casualty principle for the 
gravitational field in the effective Riemann space should be fulfilled: 
the motion of the matter subjected to the gravitational field
should not leave the light-cone limits in the Minkowsky space. 
This condition can be formulated in the form
of $g_{\mu\nu}V^\mu V^\nu\leq 0$ for any isotropic vector $V^\mu$ on the
light cone $\gamma_{\mu\nu}V^\mu V^\nu=0$. This condition being applied
to interval (15) taking into account (16) and (17) leads to 
$R^2(R^4-a)\leq 0$. Thus the constant ``$a$'' has the notion of the
fourth power of maximal value of the scale factor: 
$a=R^4_{\mbox{\small{max}}}$, and to describe the existing Universe one should
have $a>>1$.} 
\be
\left (\frac{1}{R}\frac{dR}{d\tau}\right )^2=
\frac{8\pi G}{3}\rho -\frac{\omega}{R^6}
\left (1-\frac{3R^4}{a}+2R^6\right ),
\ee
\be
\frac{1}{R}\cdot \frac{d^2R}{d\tau^2}=
-\frac{4\pi G}{3}\left(\rho +\frac{3p}{c^2}\right )
-2\omega
\left (1-\frac{1}{R^6}\right )\;,
\ee
where
\be
\omega=\frac{1}{12}\left (
\frac{m_gc^2}{\hbar}\right )^2.
\ee

It follows from Eq. (19) in the region $R>>1$ that the density of the
matter in the Universe is equal to
\be
\rho(\tau)=\rho_c(\tau)+\frac{1}{16\pi G}
\left (\frac{m_g c^2}{\hbar}\right )^2, 
\ee
where $\rho_c(\tau)$ is the critical density determined by the
Hubble ``constant'':
\be
\rho_c=\frac{3H^2(\tau)}{8\pi G},\;\;\;
H(\tau)=\frac{1}{R}\cdot \frac{dR}{d\tau}\;.
\ee
This conclusion inevitably requires the existence of the ``dark'' matter that
agrees with current observations.

From  (19) and (20) one can get the expression for the
Universe deceleration parameter $q(\tau)$.  
At the present stage of the nonrelativistic matter dominance $(p=0)$
\be
q=-\frac{\ddot R}{R}\cdot \frac{1}{H^2}=\frac{1}{2}+\frac{1}{4H^2}
\left (\frac{m_gc^2}{\hbar}\right )^2.
\ee
Relation (24) gives the principal possibility to
determine the graviton mass from two other observables, $H$ and $q$. 
The sensitivity of $q$ to the graviton mass is due to the fact that
a small value 
$\frac{1}{\lambda_g}=\frac{m_gc}{\hbar}$  comes into (24) multiplied by
a large value $\left (\frac{c}{H}\right )=9.25\cdot 10^{27}\cdot h^{-1}$~cm,
which is the Hubble radius of the Universe. Though the $q$ value has not 
been measured with high accuracy, its possible values do not exceed 
few units ($q\leq 5$, see. [4]). This allows one to estimate from 
(24) the graviton mass
\be
m_g\leq 1.7\cdot 10^{-65}\cdot h \mbox{ (g.)},\; \mbox{ where} 0.4\leq h\leq 1
\ee
\be
\frac{\hbar}{m_gc}>0.2\cdot\frac{c}{H}=2\cdot 10^{27}\cdot h^{-1}\mbox{ (cm.)}.
\ee
Despite of the smallness of the upper limit in (25), nonzero
graviton mass can have principal influence on the character of the
Universe evolution. One can see from Eq. (19) that for $R\to 0$ 
the negative term $\frac{\omega}{R^6}$ in the right-hand side
of the equation grows in absolute value faster than the matter density
($\rho\sim\frac{1}{R^4}$ for radiatively dominant stage). Therefore, from
the condition of the nonnegative left-hand side of (19) it follows
that the expansion should begin from some minimal value
$R_{\mbox{\small{min}}}$, which corresponds to $\frac{dR}{d\tau}=0$.
From other side, the expansion should stop at $R>>1$, when
density (22) reaches its minimal value 
$\rho_{\mbox{\small{min}}}=\frac{1}{16\pi G}
\left (\frac{m_gc^2}{\hbar}\right )^2$, and after that the expansion
is replaced by the compression process up to $R_{\mbox{\small{min}}}$. 
So, nonzero graviton mass eliminates the cosmological singularity
and leads to cyclic character of the Universe evolution.
Such a character of the Universe evolution seemed to be promising 
for a number of authors  (see, for example, [5]). 
The time of the Universe expansion from the maximal density to the minimal
one is determined mostly by the stage of the nonrelativistic matter
dominance, and it is equal to  [6]
\be
\tau_{\mbox{\small{max}}}\simeq\sqrt{\frac{2}{3}}\cdot 
\frac{\pi\hbar}{m_gc^2}\;.
\ee  
Accepting the value of $(10-15)\cdot 10^9$ years for the Universe age
and using $\tau_{\mbox{\small{max}}}\geq 20\cdot 10^9$ years,
on gets more strict limit on the graviton mass
\be
m_g\leq 4.5\cdot 10^{-66}\mbox{ (g.)}\;.
\ee
Equations (12) and (13) explain all known gravitational effects in the Solar 
system, which are attributed to the noninertial frame. It is well known
that the introduction of the graviton mass in the linear tensor theory 
is accompanied by the difficulty: the presence of ``ghosts''. 
However, as it was shown in Ref. [7], this difficulty is eliminated
in the framework of the nonlinear tensor theory described by
Eqs. (12) and (13) under condition that gravitons spread in the effective 
Riemann space, rather than Minkowsky's one (as it takes place in the 
linear theory). If this circumstance is sequently taken into consideration,
one gets a positively determined flux of the gravitational energy when 
calculating the intensity.

\end{document}